\documentclass[10pt,a4paper]{article}

\usepackage[utf8x]{inputenc}
\usepackage{ucs}
\usepackage{amsmath,amsthm, url}
\usepackage{amsfonts}
\usepackage{amssymb}
\usepackage{geometry,color,graphicx,url}
 \usepackage{multirow}
\usepackage[english]{babel}
\usepackage{setspace}
\usepackage{hyperref}
\usepackage{graphicx,bbm}
\usepackage{graphicx}
\usepackage{subfigure}
\usepackage[toc,page]{appendix} 
\usepackage{csquotes}

\title{Study of the Parent-of-origin effect in monogenic diseases with variable age of onset.\\ Application on ATTRv }

\author{Flora Alarcon$^1$, Violaine Plant\'e-Bordeneuve$^{2}$ , Gregory Nuel$^3$}

\date{}

\begin{document}
\doublespacing
\maketitle

\hspace{-0.65cm}
$^1$ Laboratory MAP5 UMR CNRS 8145 Paris University, Paris, France\\
$^2$ Department of Neurology, Henri Mondor University Hospital, APHP, Créteil, France. And Paris Est-Créteil University, Créteil, France. And Inserm U.955, Institut Mondor de Recherche Biomédicale (IMRB), Créteil, France. \\
$^3$ Stochastics and Biology Group, Department of Probability and Statistics (LPSM, UMR CNRS 8001), Sorbonne University, Paris, France\\


\section*{Abstract:}

In genetic diseases with variable age of onset, an accurate estimation of the survival function for the mutation carriers and also modifying factors effects estimations are important for the management of asymptomatic gene carriers across life.
Among the modifying factors, the gender of the parent transmitting the mutation (i.e. the parent-of-origin effect) has been shown to have a significant effect on survival curve estimation on transthyretin familial amyloid polyneuropathy (ATTRv) families. 
However, as most genotypes are unknown, the parent-of-origin must be calculated through a probability estimated from the pedigree.
We propose in this article to extend the method providing mutation carrier survival estimates in order to estimate the parent-of-origin effect.
The method is both validated on simulated data and applied to familly samples with ATTRv.

\vspace{1cm}
\noindent\textbf{Key words:Semi-parametric estimation, Survival Curve, Parent of origin effect, Proportionnal hazard model, Familial amyloid polyneuropathy.}

\section{Introduction}
In variable age of onset diseases caused by a deleterious gene mutation, knowledge of the survival function for carrier individual is important both to understand the underlying mechanism of the disease (like identification of potential factors that modulate this age of onset) and for prevention strategies. 
In the literature on this subject, the age-specific cumulative distribution function (CDF), named also penetrance function, is preferentially used. In this paper, like in \cite{alarcon2018non},  we will use the classical survival function (which is simply the complementary of the CDF) to assess the probability of not being affected by the disease according to the age for mutation carrier individuals. Note that our survival function hence corresponds to the \emph{cause-specific} survival (disease diagnosis) and not to the \emph{overall} survival.

Recently, a semi-parametric method based on a Cox model was developed to estimate survival function from familial data ascertained through affected individual. This method is very efficient and handle the unknown genotypes through the sum-product algorithm in Bayesian network. The method is described in detail in \cite{alarcon2018non}. 
This method has been applied on families affected by a mendelian disease : the ATTRv, that is the most frequent familial amyloidosis, with autosomal dominant transmission. A fatal outcome occurs after an average duration of 10-13 years \cite{plante2000transthyretin, suhr1994malnutrition}. This severe diseases shows important differences in age of onset and thus on survival curve, according to different covariates as countries, gender, mutations \cite{hellman2008heterogeneity,Holmgren,gorram2020new}. Especially, heterogeneity in the survival curves according to gender of the transmitting parent have been noted on Portugueses ATTRv families \cite{Sousa,ines2018epidemiology}. In particular, Hellman et al. \cite{hellman2008heterogeneity} have find that the risk of disease in the carriers was significantly higher when the mutation was inherited from the mother than from the father. 

In 2009, Bonaïti et al \cite{bonaiti2009parent} investigate the parent-of-origin effect in a sample of French and Portuguese families that have already been described in \cite{plantebordeneuve2003gst}. The covariate for parent of origin factor was not calculated algorithmically according to pedigrees and genotypical information but was determined previously to the estimation of the parameters and decided manually by expert-based review.
They found  that the penetrance was higher when the mutation was inherited from the mother than from the father (i.e. the survival curve was lower when the mutation was inherited from the mother than from the father). This difference was significant in the Portuguese families but not in the French sample. This results led them to the hypothesis of a genetically determined effect through an imprinting phenomenon.

Unfortunately, the gender of the transmitting parent is not known for most of carriers du to the fact that most of individuals are non-genotyped. And we should be skeptical as to the estimates made in this context. However, no simulation studies have been proposed to evaluate the quality of survival curve estimations in studying the parent-of-origin effect.

In this article, we propose to extend the semi-parametric method based on a Cox model described in \cite{alarcon2018non} in order to determine the gender of the transmitting parent. 
The method for analysing the parent-of-origin effect in survival curve estimations of Mendelian disease when a part of genotypic information is unknown, is validated through simulations.

To do the link with the literature about the subject, we illustrate our results on Portugese fand French datasets of ATTRv already analyzed in \cite{bonaiti2009parent} and compare our results with theirs.


\section{Methods}
Survival curve are estimated with a semi-parametric method based on a Cox model and adapted for pedigree data. As mentioned before, the data available are families ascertained through at least one affected individual. For all individuals, we have information about his genotype, phenotype, age, gender, etc.
Families data consist of individuals independents from each other conditionally to their genotype. The notion of family is taken into account exclusively in the estimation of the probability to carry the mutation. 
The method is described in detail in \cite{alarcon2018non}. In this section, the method is briefly presented in the particular case of studying the parent-of-origin effect.


\subsection*{The model}

If we denote by $\mathrm{ev}$ the \emph{evidence} which consists of all the available information: all time $T_i$ and status $\delta_i$, the individual genotype $X_i$ and covariates $Z_i$ (possibly multidimensional: gender, comorbidities, ethnicity, etc.), and the genetic testing $G_i$ (when available, $\mathrm{NA}$ is missing) we have:
$$
\mathbb{P}(X,\mathrm{ev})=\prod_i \underbrace{ \mathbb{P}(T_i,\delta_i | X_i, Z_i) \mathbb{P}(G_i | X_i)^{\mathbf{1}_{G_i \neq \mathrm{NA}}}}_{\phi_i\left(X_i\right)} \prod_{i \in \mathcal{F}} \mathbb{P}\left(X_i\right) \prod_{i \notin \mathcal{F}}
\mathbb{P}\left(X_i | X_{\mathrm{pat}_i}, X_{\mathrm{mat}_i}\right)
$$
where $\mathcal{F}$ is the set of founders and where $X_{\mathrm{pat}_i}$ and $X_{\mathrm{mat}_i}$ indicate respectively the genotype of the father and the mother on individual $i$. Thus, the genotype distribution among founders follows the Hardy-Weinberg equilibrium with disease allele frequency $q$, and the conditional distribution for non-founders follows the Mendelian transmission of alleles. If $G_i = \mathrm{NA}$, then $X_i \in \{0, 1_m, 1_p, 2 \}$ corresponding respectively to a non-mutated individual, a mutated individual whose mutation comes from the father, a mutated individual whose mutation comes from the mother and finally, to a homozygous individual.

More precisely, we have : 
$$
\mathbb{P}(T_i=t,\delta_i=0 | X_i=x, Z_i)= \left\{
\begin{array}{ll}
\exp\left( -\Lambda_0(t) \exp(Z_i\gamma) \right) & \text{if $x=1\mathrm{m}$ or $x=2$}\\
\exp\left( -\Lambda_0(t) \exp(\beta + Z_i\gamma) \right) & \text{if $x=1\mathrm{p}$}\\
1 & \text{if $X_i=0$}
\end{array}
\right.
$$
and
$$
\mathbb{P}(T_i=t,\delta_i=1 | X_i=x, Z_i)= \left\{
\begin{array}{ll}
\exp\left( -\Lambda_0(t) \exp(Z_i\gamma) \right) \times \left( \lambda_0(t) \exp(Z_i\gamma) \right) & \text{if $x=1\mathrm{m}$ or $x=2$}\\
\exp\left( -\Lambda_0(t) \exp(\beta + Z_i\gamma) \right)\times \left( \lambda_0(t) \exp(\beta + Z_i\gamma)\right) & \text{if $x=1\mathrm{p}$}\\
0 & \text{if $X_i=0$}
\end{array}
\right.
$$

which can be simplified as follows : 

$$
\frac{1}{\lambda_0(t)} \mathbb{P}(T_i=t,\delta_i=1 | X_i=x, Z_i)= \left\{
\begin{array}{ll}
\exp\left( -\Lambda_0(t) \exp(Z_i\gamma) \right)\exp(Z_i\gamma) & \text{if $x=1\mathrm{m}$ or $x=2$}\\
\exp\left( -\Lambda_0(t) \exp(\beta + Z_i\gamma) \right)\exp(\beta + Z_i\gamma) & \text{if $x=1\mathrm{p}$}\\
0 & \text{if $X_i=0$}
\end{array}
\right.
$$
where $\lambda_0$ is the baseline hazard and $\Lambda_0$ the baseline cumulative hazard. Moreover, $\beta$ is the Cox's model parameter, to be estimated, for the parent-of-origin effect and $\gamma$ the parameter vector corresponding to the other covariates.
For the genetic testings, we have a simple model with false positive rate $\varepsilon$ and false negative rate $\eta$:
$$
\mathbb{P}(G_i=1 | X_i \neq 0)=1-\varepsilon
\quad
\mathbb{P}(G_i=0 | X_i = 0)=1-\eta
$$

Note that, in our case, these rates are probably very small (e.g. $\varepsilon < 1/100$ and $\eta < 1/1000$)

\subsection*{EM framework}
As $X_i$ is either partially observed or not observed, we consider this variable as latent and use a classical Expectation-Maximization algorithm in order to maximize the log-likelihood model in parameter of interest.
 
Model parameters are: $q$, $\beta$, $\gamma$, $\varepsilon$, $\eta$, $\Lambda_0$ and $\lambda_0$. In order for the model to be identifiable, we assume that error rates $\varepsilon$ and $\eta$ are known as well as the disease allele frequency $q$. And since $\lambda_0$ appears only has a proportional factor in the expression of $\phi_i(X_i)$ we can perform model inference without explicit value for this parameter. Our aim is therefore to estimate $\theta=(\beta,\gamma,\Lambda_0)$ using a classical EM framework.

In this framework we alternate to steps until convergence:
\begin{description}
\item[E-Step] give the weights $w_i$ using current parameter $\theta_\text{old}$, computed for all $i$:
$$
w_i^\mathrm{pat}=\mathbb{P}\left(X_i=1\mathrm{p} | \mathrm{ev} ; \theta_\text{old}\right)
\quad\text{and}\quad
w_i^\mathrm{mat}=\mathbb{P}\left(X_i=1\mathrm{m} \text{ or } X_i=2  | \mathrm{ev}; \theta_\text{old}\right)
$$
\item[M-Step]: create an artificial weighted dataset with the following $2n$ patients 
$$
\begin{array}{ccccc}
\hline
\text{time} & \text{status} & \text{POO} & \text{covariates} & \text{weight} \\
\hline
T_1&\delta_1&\mathrm{pat}& Z_1 & w_1^\mathrm{pat} \\
\vdots&\vdots&\vdots& \vdots& \vdots \\
T_n&\delta_n&\mathrm{pat}& Z_n & w_n^\mathrm{pat} \\
\hline
T_1&\delta_1&\mathrm{mat}& Z_1 & w_1^\mathrm{mat} \\
\vdots&\vdots&\vdots& \vdots& \vdots \\
T_n&\delta_n&\mathrm{mat}& Z_n & w_n^\mathrm{mat} \\
\hline
\end{array}
$$
and simply:
\begin{itemize}
\item fit a (weighted) Cox's proportional hazard model with Factor POO and the covariates to update $\beta$ and $\gamma$
\item use non-parametric estimate (\emph{e.g.} Kaplan-Meier or similar) of $\Lambda_0$.
\end{itemize}
\end{description}

\subsection*{Posterior distributions}

For computing the posterior weights $w_i^\mathrm{pat}$ and $w_i^\mathrm{mat}$ of the E-Step one has to integrate the model over all unobserved (even when partially observed) genotypes $X_i$. Since the total number of configurations for $X_i$ is $4^n$ it is clearly impossible to perform this summation with brute force. Fortunately, geneticist know for year that likelihood computations in genetic model can be performed efficiently using the Elston-Stewart algorithm\cite{elston1971general}. Thanks to this algorithm, it is therefore possible to compute any posterior distribution as a simple likelihood ratio:
$$
\mathbb{P}(X_i=x | \mathrm{ev}) = \frac{\mathbb{P}(X_i=x, \mathrm{ev})}{\mathbb{P}(\mathrm{ev})}
$$
In the probabilistic graphical model community, such computation can be dealt efficiently using the sum-product algorithm (also called: belief propagation, forward/backward, inward/outward, Kalman filter, etc.) in order to obtain in a single computational pass \emph{all}
$w_i^\mathrm{pat}$ and $w_i^\mathrm{mat}$. NB: in Totir \cite{totir2009efficient} , the authors suggest a generalization of the Elston-Stewart algorithm allowing to compute all posterior distribution in a single passe rather than repeating likelhood computations. Without surprise, Totir's algorithm is in fact the exact reformulation of the classical sum-product algorithm.

Practically, initialization is performed by affecting random weights $w_i$ (ex: drawn from a uniform distribution on $[0,1]$). EM iterations are stopped when we observe convergence on test survival estimates (ex: baseline survival at age $20,40,60,80$). 

\section{Results}
\subsection*{Simulations study}

\begin{figure}[htbp]
    \begin{center}
        \includegraphics[trim=10 200 10 250,clip,width=0.7\textwidth]{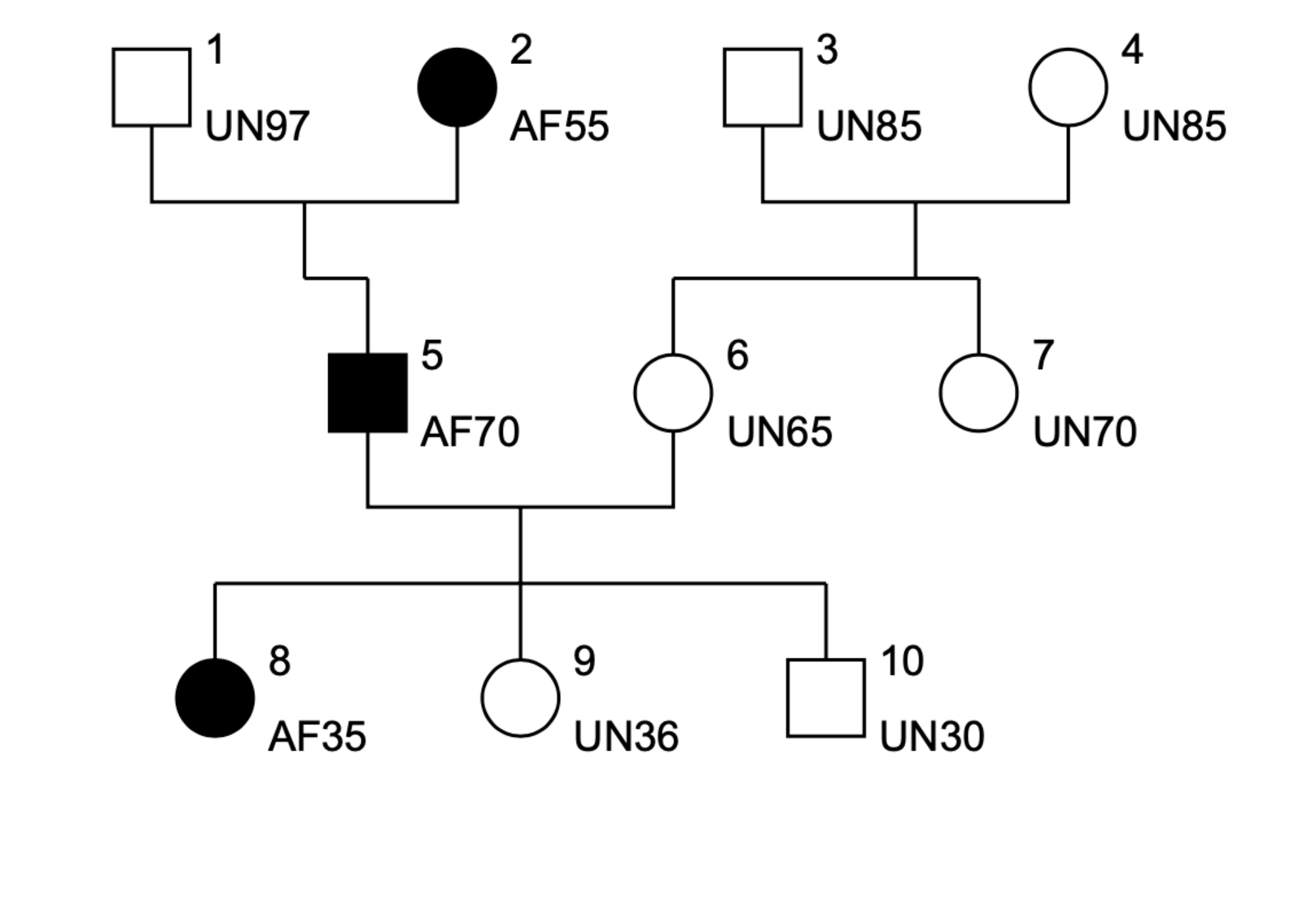}
    \caption{Simulated family structure}
    \label{Pedigree}
    \end{center}
\end{figure}

We have simulated $n$ families ($n=100$ or $400$) with 10 individuals  as shown in figure \ref{Pedigree}. 

Genotypes were assigned respecting Mendelian transmission and heterozygous genotypes were ordered according to the gender (pat or mat) of the parent who transmitted the mutation. 
So, a new binary covariate (named POO for parent of origin) are introduced such as $\text{POO} = \mathrm{mat}$ if the mother transmitted the disease mutation and $\text{POO} = \mathrm{pat}$ if disease mutation was transmitted by the father. 
The disease allele frequency was set to q = 0.20 in our simulated dataset for the sake of speed (without simulating any ascertainment process). 
The age at event was simulated according to a piecewise constant hazard rate function. Moreover a censoring variable was added that follows a uniform distribution $\mathcal{U}[15,80]$. 

To study the parent-of-origin effect,  the POO was simulated in a proportional hasard model. The risk when the mutation is transmitted by the mother is given by $\lambda_{0}$ as follows and the coefficient parameters in the hazard model is noted $\beta$ ($\beta = -0.60$ or $\beta=-1.2$ in simulations). Thus, as $\beta < 0$, that's mean that survival is upper when mutation provides on the father.

$$
\lambda_{0}(t)=\left\{
\begin{array}{ll}
0 & \text{ if } t \in [0, 20]\\
0.02 & \text{ if } t \in [20, 40]\\
0.10 & \text{ if } t \in [40, 60]\\
0.05 & \text{ if } t > 60 
\end{array}
\right.
$$

To study the behavior of our method, we will consider different scenarios for simulations. In the first scenario (S0), all genotypes are set to unobserved, in the second more realistic scenario (S1), 80\% of genotypes are observed in affected individuals and only 10\% are observed in non affected individuals and in the third one (S2), all genotypes are observed. Moreover, we consider an other scenario where, in addition to the genotypes, we know, for each individual, the sex of the parent transmitting the mutation. We call this scenario \enquote{Oracle}. 
In order to compare the $\beta$ estimate in the different scenarios, 200 replications are performed.

\begin{figure}[htp!]
\centering
\includegraphics[height=0.8\textwidth]{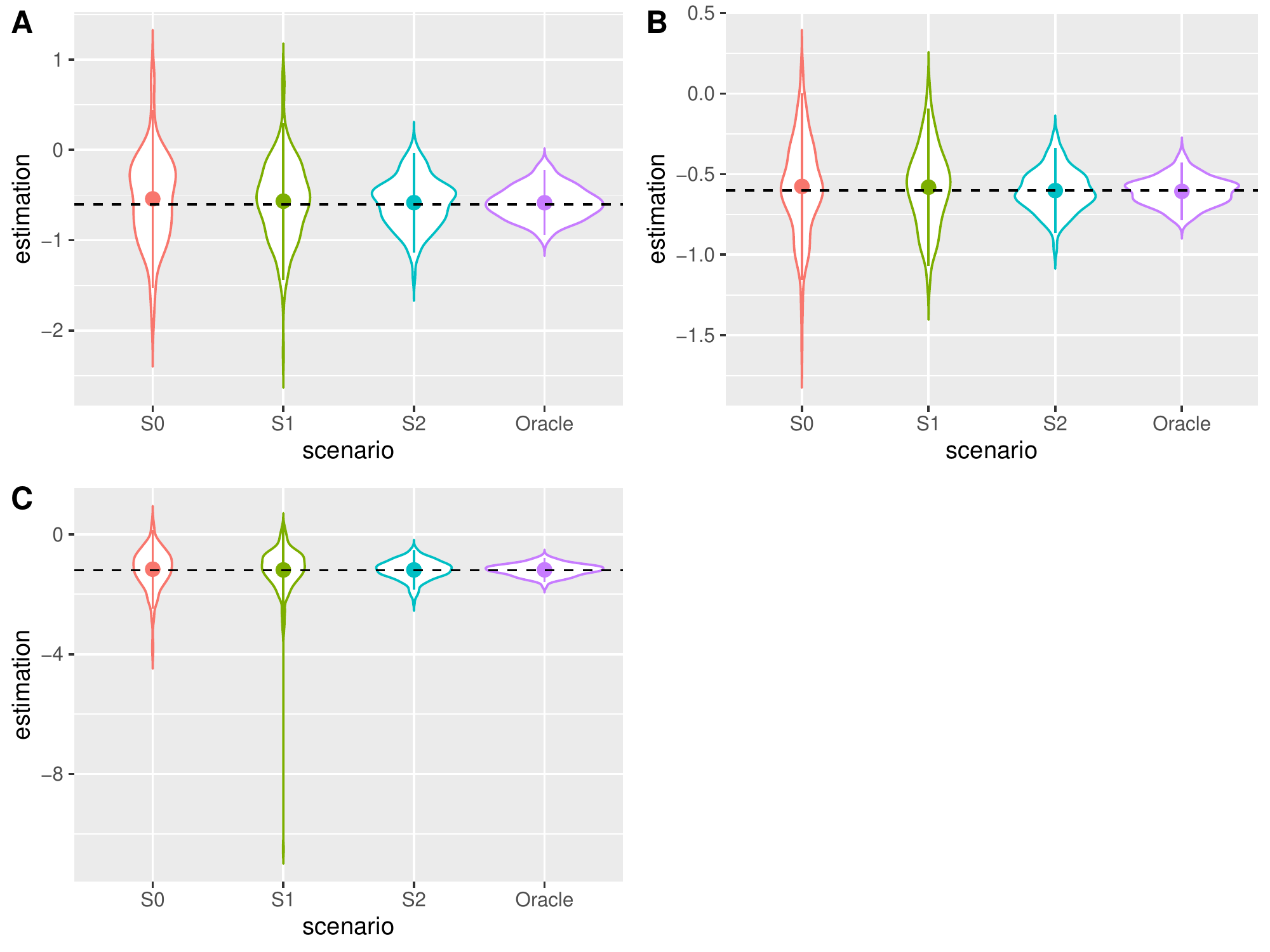}
\caption{Violin Plots according to different scenarios. \textbf{\textsc{A}} : $n=100$ families ; $\beta = -0.6$. \textbf{\textsc{B}} : $n=400$ families ; $\beta = -0.6$. \textbf{\textsc{C}} : $n=100$ families ; $\beta = -1.2$}\label{violinplots}
\end{figure}
\begin{figure}[htp!]
\centering
\includegraphics[height=0.6\textwidth]{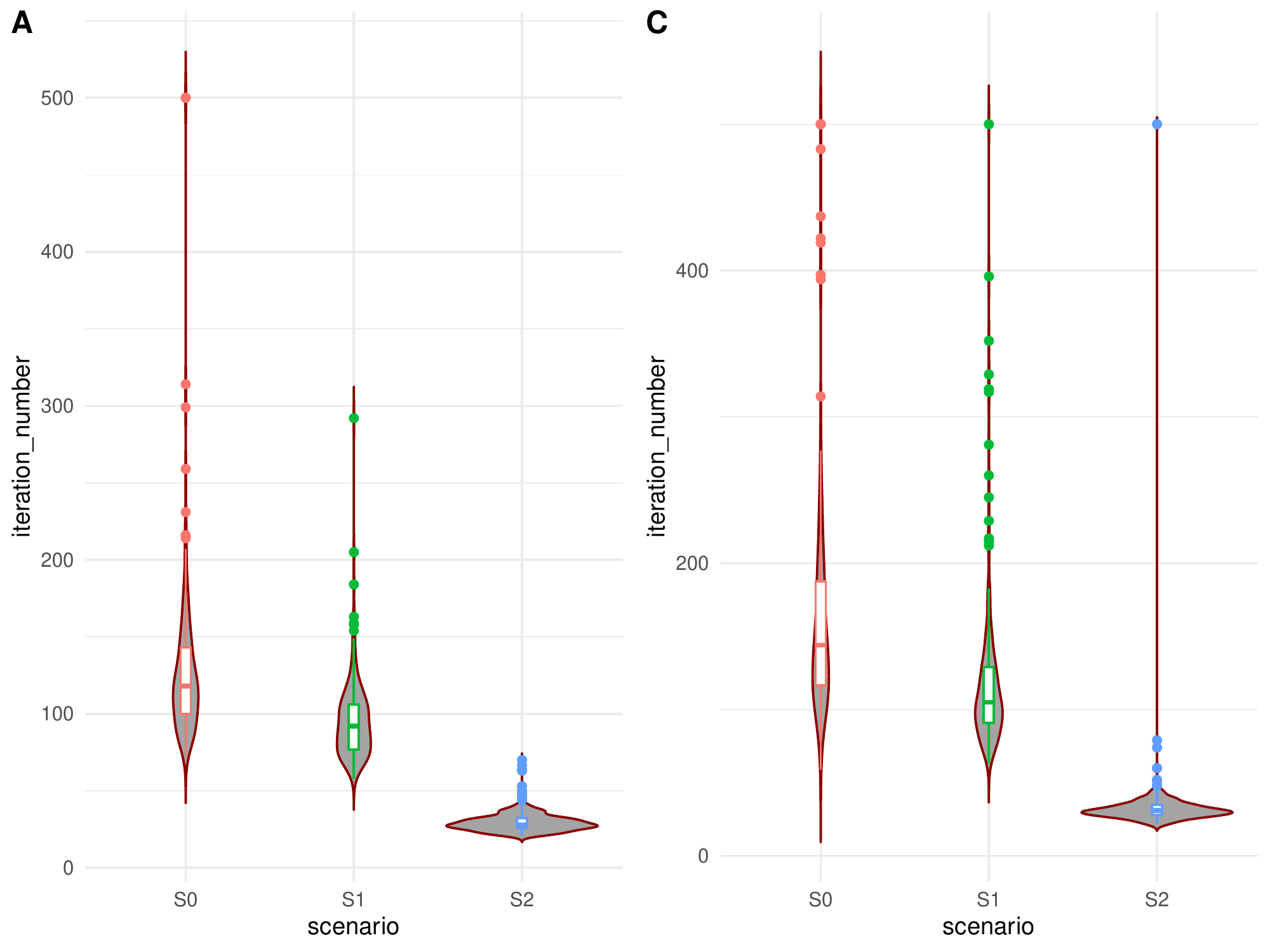}
\caption{Violin plot with box plot of the number of iterations before EM convergence according to different scenarios. \textbf{\textsc{A}} : $n=100$ families ; $\beta = -0.6$. \textbf{\textsc{C}} : $n=100$ families ; $\beta = -1.2$}\label{violinplots-iter}
\end{figure}

Figure \ref{violinplots} shows the violin plots according to the scenario for the percentage of observed genotypes, which allows to show the full distribution of the data. Thus, each color corresponds to a genotypic  scenario for each of the cases \textbf{\textsc{A}} ($n=100$ families was simulated with $\beta=-0.6$), \textbf{\textsc{B}} ($n=400$ families was simulated with $\beta=-0.6$) and \textbf{\textsc{C}} ($n=100$ families was simulated with $\beta=-1.2$).
For each scenario, the mean and standard deviation are shown and the horizontal black line represents the true $\beta$ value. 
In all cases the $\beta$ estimate are unbiased whatever the scenario and the standard deviation increases with the percentage of unobserved genotypes. As expected, the oracle varies little. 
It is also interesting to look at the number of iterations needed before convergence of the EM algorithm, depending on the different scenarios. 
Figure \ref{violinplots-iter} shows the violin plot including boxplots of the number of iterations required before the convergence of the EM algorithm according to scenarios (S0), (S1) and (S2) in the cases \textbf{\textsc{A}} and \textbf{\textsc{B}}. The violin plot shows that this number increases with the amount of unknown genotypes. Thus, when all genotypes are observed, the algorithm converges in 65 iterations on average while it converges in 89 iterations on average when no genotypes are observed. We also observe a higher variability of the number of iterations when the genotypes are not observed.


\subsection*{Application}


%
%

\begin{figure}[htp!]
\centering
\includegraphics[height=0.6\textwidth]{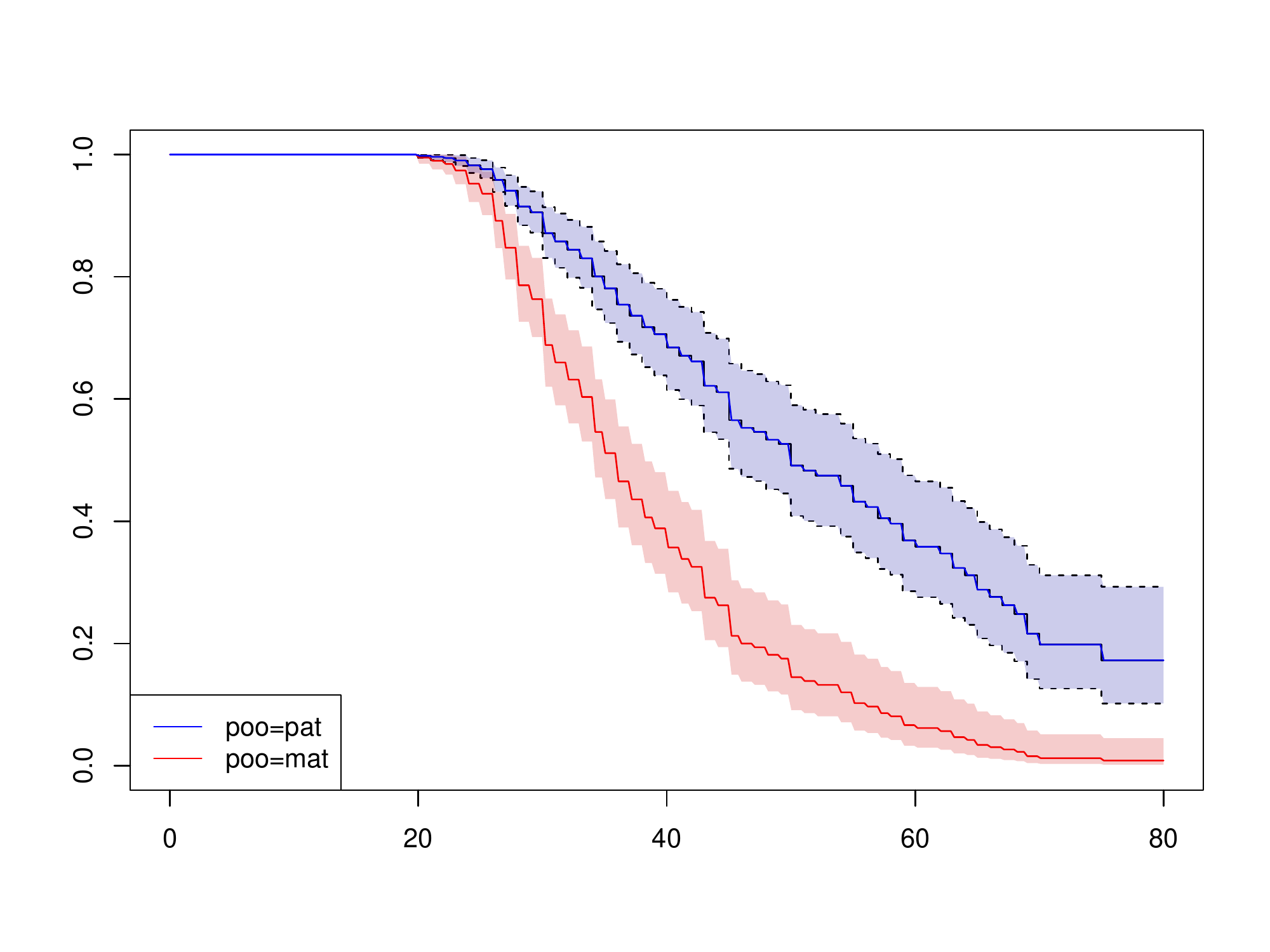}
\caption{Estimation of the survival function according to the sex of the transmitting parent with 95\% confidence intervals for Portugese families}\label{Cox-poo-port}
\end{figure}

Our method has been applied to two datasets of Portuguese families and French families already analyzed in \cite{bonaiti2009parent}. For data analysis, the disease allele frequency was set to $q = 0.04$ as in \cite{hellman2008heterogeneity}.
The ascertainment bias was corrected by a classical method that consists in simply removing the phenotypic information of the proband, as done in \cite{alarcon2018non}.
The parent of origin gender parameter was not significantly different from zero in French families ($\hat{\beta} = 0.114$ with p-value $=0.467$). 
However, the $\beta$ estimate was significantly different from zero in Portuguese families ($\hat{\beta} = -0.999$ and a p-value $ = 5. 10^{-10}$) showing that the risk of being affected is higher when the mutation is transmitted by the mother than by the father. This difference in survival curves estimated is shown in Figure \ref{Cox-poo-port} and is totally consistent with the results of \cite{bonaiti2009parent}.

\section{Conclusion}
In this article, we propose an extension of our method in order to take into account the gender of the transmitting parent in the estimation of the survival function from familial data in cases of age-dependent genetic diseases. The probability of the transmitting parent gender is calculated and its effect is estimated through a proportional hazard model. Our extension is assessed with simulations and applied on a French and a Portuguese dataset of ATTRv families. Obviously, the method provides confidence intervals for each estimation.

The simulation study shows that the method is unbiased and the increase in variance of the estimate with the number of unobserved genotypes.

We have applied this method to a French and Portuguese dataset and shows a significant difference of the penetrance function according to the gender of the transmitted parent. 
These results are consistent with the analyses already done in \cite{bonaiti2009parent}, even if the method used in \cite{bonaiti2009parent} was less precise since the genotypic probabilities were not calculated. Here we have proposed a unified method able to take into account precisely these probabilities. 
Moreovern the methode allowsi additional variables to be taken into account when estimating survival curves.

Finally, in this work, genotypic probabilities are calculated using a sum-product algorithm which is a very general method which can deal efficiently with very complex pedigree structure (ex: $2000$ individuals with $50$ loops). Unlike Elston-Stewart algorithm, the sum-product algorithm does not use loop breaking approaches to deal with loop pedigrees. Instead, the sum-product algorithm use an auxiliary tree called the junction tree (JT) which basically is a clique decomposition of the moral graph corresponding to the pedigree problem. JT and BP are well known is the graph theory (ex: JT can be used to solve a graph coloring problem) and in the mathematical field of probabilistic graphical models (Bayesian network, hidden Markov model, decision trees, Markov networks, etc.).


\newpage

\noindent \textbf{Conflict of interest.} The authors declare no conflict of interest.

\bibliographystyle{unsrt}
\bibliography{biblio-NAH-POO}

\end{document}